\documentclass[fleqn]{article}

\usepackage{amsmath,amssymb}
\usepackage{graphicx}
\usepackage[top=0.75in, bottom=0.75in, left=0.75in, right=0.75in, dvips]{geometry}

\begin{document}

\title{{\sf A Unique Determination of the Effective Potential in Terms of Renormalization Group Functions}}
\author{
F.A.\ Chishtie\thanks{Department of Applied Mathematics, The
University of Western Ontario, London, ON  N6A 5B7, Canada}, 
T.\ Hanif\thanks{Department of Applied
Mathematics, The University of Western Ontario, London, ON N6A
5B7, Canada},
D.G.C.\ McKeon\thanks{Department of Applied
Mathematics, The University of Western Ontario, London, ON N6A
5B7, Canada}, T.G.\ Steele\thanks{Department of Physics and
Engineering Physics, University of Saskatchewan, Saskatoon, SK,
S7N 5E2, Canada} }

\maketitle

\maketitle

\begin{abstract}
The perturbative effective potential $V$ in the
massless $\lambda\phi^4$ model with a global $O(N)$ symmetry is
uniquely determined to all orders by the renormalization group functions alone
when the Coleman-Weinberg renormalization condition
$\left.\frac{d^4V}{d\phi^4}\right|_{\phi = \mu} = \lambda$ is
used, where $\mu$ represents the renormalization scale. 
Systematic methods are developed to express the $n$-loop effective potential in the Coleman-Weinberg scheme
in terms of the known $n$-loop minimal subtraction  (MS) renormalization-group functions. 
Moreover, it also proves possible to sum the leading- and subsequent-to-leading-logarithm
contributions to $V$.
An
essential element of this analysis is a conversion of the renormalization group functions in the Coleman-Weinberg
scheme to the renormalization group functions in the MS scheme.
As an example, the explicit five-loop effective potential is obtained from the known five-loop MS renormalization group functions and we explicitly sum the  leading logarithm ($LL$), next-to-leading ($NLL$), and further 
subleading-logarithm contributions to $V$.
Extensions of these results to massless scalar QED are also presented.  Because massless scalar QED has two couplings, conversion of the RG functions from the MS scheme to the CW scheme requires the use of multi-scale renormalization group methods.
\end{abstract}

\section{Introduction}
The effective potential $V$ in the massless $\lambda\phi^4$
model with a global $O(N)$ symmetry has received considerable
attention \cite{1,2,3,4,5} because of its connection to the
scalar-field theory projection of the Standard Model for $N=4$. A
variety of renormalization schemes have been employed in computing
the effective potential, among them minimal subtraction (MS) \cite{6} and
imposing the Coleman-Weinberg (CW) condition \cite{1,3,7}
\begin{equation}
\left.\frac{d^4V}{d\phi^4}\right|_{\phi = \mu} = 24 \lambda~,
\label{eq1}
\end{equation}
where $\mu$ is the renormalization scale.

In this paper we develop iterative techniques that uniquely determine leading-logarithm and subsequent-to-leading-logarithm expansions of the effective potential in the CW scheme for  $O(N)$-symmetric massless $\lambda\phi^4$ theory and for massless scalar QED with a $\phi^4$ interaction. As  discussed below, the renormalization-group (RG) functions in the CW scheme differ from the known minimal-subtraction (MS) scheme RG functions \cite{7}, resulting in non-trivial effects of this scheme conversion beginning at two-loop order.  Although it has been known for some time that the effective potential  is in principle determined by the RG equation \cite{1}, two-loop calculations have either failed to make the necessary scheme conversion  \cite{10,ndili}
or have been done explicitly without using RG methods \cite{3}. 

In Section \ref{phi_sec} we explicitly construct the effective potential $V$ for the $\lambda \phi^4$ model, not only up to five loop order, but also give all the $N^4LL$ (next-to, next-to, next-to, next-to leading-logarithm) contributions to $V$ without any explicit evaluation of any diagrams, simply by applying the RG equation in conjunction with the CW renormalization scheme, thereby realizing the result of Ref.~\cite{1}.\footnote{The relation between $V$ and the RG equation appearing in Ref.~\cite{1} is further analyzed in Ref.~\cite{vic}  where the importance of fixing the constants $T_{n0}$ appearing in Eq.~(\protect\ref{eq2}) is emphasized.}   

In Section \ref{sqed_sec} we extend our analysis to massless scalar QED with a $\phi^4$ interaction, a theory which contains two couplings.  The results are quite similar to the $\lambda\phi^4$ scenario; iterative methods are developed to determine the scalar-field effective potential in terms of the RG functions in the CW scheme.  The presence of multiple couplings requires the use of multi-scale RG methods \cite{multi_scale_einhorn,multi_scale_ford} to convert the RG coefficients to the CW scheme.

\section{Massless $O(N)$-Symmetric $\lambda\phi^4$  Theory}\label{phi_sec}
In $O(N)$-symmetric massless $\lambda\phi^4$  theory, the effective potential in the CW scheme takes the form
\begin{equation}
V(\lambda, \phi, \mu) = \sum^\infty_{n=0} \sum^n_{m=0} \lambda^{n+1} T_{nm}L^m\phi^4
\label{eq2}
\end{equation}
when computed in perturbation theory, where we have defined
\begin{equation}
L = \log \left(\frac{\phi^2}{\mu^2}\right)~.
\label{eq3}
\end{equation}
Since the renormalization scale $\mu$ is unphysical, changes in
$\mu$ must be compensated for by changes in $\lambda$ and $\phi$;
this leads to the renormalization group equation
\begin{equation}
\left(\mu \frac{\partial}{\partial \mu} + \beta(\lambda) \frac{\partial}{\partial \lambda} +
\gamma(\lambda) \phi \frac{\partial}{\partial \phi} \right) V(\lambda, \phi , \mu) = 0
\label{eq4}
\end{equation}
where
\begin{gather}
\beta(\lambda) = \mu \frac{d\lambda}{d \mu} = \sum^\infty_{k=2} b_k \lambda^k
\label{eq5}
\\
\gamma(\lambda) = \frac{\mu}{\phi} \frac{d \phi}{d\mu} = \sum^\infty_{k=1} g_k \lambda^k ~.
\label{eq6}
\end{gather}

The renormalization group equation (\ref{eq4}) can be used \cite{5,8} to sequentially sum the logarithms appearing in 
Eq.~(\ref{eq2}); that is if we rewrite Eq.~(\ref{eq2}) as
\begin{equation}
V (\lambda , \phi, \mu) = \sum^\infty_{n=0} \lambda^{n+1} S_n (\lambda L)\,\phi^4
\label{eq7}
\end{equation}
where
\begin{equation}
S_n (\lambda L) = \sum^\infty_{m=0} T_{n+m\, m}(\lambda L)^m~,
\label{eq8}
\end{equation}
then the functions $S_n(\xi)$ are determined by Eq.\  (\ref{eq4}) provided we impose the boundary condition $S_n(0) = T_{n0}$. More explicitly, we write Eqs.\ (\ref{eq4}--\ref{eq7}) as
\begin{equation}
\begin{split}
0=\sum^{\infty}_{k=0} &
\left[-1 + \left(g_1 \lambda + g_2\lambda^2 +\ldots\right)\right]
2\lambda^{k+2}S_k^\prime(\xi)
+ \left( b_2\lambda^2 + b_3\lambda^3 + \ldots\right)\lambda^k
\left[ (k+1) S_k (\xi)
+ \xi S_k^\prime (\xi)\right]
\\
& + 4 \left( g_1\lambda + g_2 \lambda^2 + \ldots\right) \lambda^{k+1} S_k (\xi)~,
\end{split}
\label{eq9}
\end{equation}
so that to order $\lambda^2$, and in general to order $\lambda^{n-2}$,  we respectively find that
\begin{gather}
\left[\left(-2 + b_2 \xi\right)\frac{d}{d\xi} + \left(b_2 +4g_1\right)\right] S_0 = 0
\label{eq10}
\\
\begin{split}
0=&\left[
\left(-2+b_2\xi\right)\frac{d}{d\xi}+\left(n+1\right)b_2+4g_1
\right] S_n
\\
&+
\sum_{m=0}^{n-1}\left\{
\left(2g_{n-m}+b_{n+2-m}\xi\right)\frac{d}{d\xi}+
\left[ \left(m+1\right)b_{n+2-m}+4g_{n+1-m}  \right]
\right\} S_m~.
\end{split}
\label{neweq11}
\end{gather}
In general, if Eq.\  (\ref{eq9}) is satisfied at order $\lambda^{n+2}$, $S_n(\xi)$ satisfies the differential equation
(\ref{neweq11})
whose solution requires that we know $S_m(\xi)$ ($m = 0,~1, ~ \ldots n-1$), $b_m$ ($m=2,~  \ldots n+2$), $g_m$
($m=1,~\ldots  n+1$), and the boundary condition $S_n(0)=T_{n0}$.
In other words, $S_n(\xi)$ is governed by coupled differential equations which depend upon the $n+1$ loop
RG coefficients. It is important to note that the RG equation by itself does not determine the boundary conditions $T_{n0}$; these will be seen to be determined by the CW renormalization condition.

The solutions for $S_0$ and $S_1$ are
\begin{gather}
S_0(\xi)=\frac{T_{00}}{w}
\label{S0_sol}
\\ 
S_1(\xi)=-\frac{4g_2T_{00}}{b_2w}+\frac{4g_2T_{00}+b_2T_{10}}{b_2w^2}
-\frac{b_3T_{00}}{b_2 w^2}\log{w}~,
\label{S1_sol}
\end{gather}
where
$w = 1 - \frac{b_2}{2} \xi$ and 
$g_1 = 0$ for the $\lambda\phi^4$ model.
As suggested by (\ref{S0_sol}) and (\ref{S1_sol}),   the explicit solutions to Eqs.\ (\ref{eq10}) and (\ref{neweq11}) take the form of polynomials in $w$ and $\log w$: 
\begin{equation}
S_n(\xi)=\frac{1}{b_2}\sum_{i=1}^{n+1}\sum_{j=0}^{i-1}\sigma^{n}_{i,j}\frac{\left[\log w\right]^j}{w^i}~.
\label{Sn_sigma_exp}
\end{equation}
Expressions for the coefficients $\sigma^{n}_{i,j}$ up to $n=2$ are given in  Appendix A in terms of a  recursion relation.  Appendix A also demonstrates that a partial summation of the recursion relation is possible.  

If $V$ is computed so that Eq.\  (\ref{eq1}) is satisfied, then upon substituting Eq.\  (\ref{eq7}) into Eq.\  (\ref{eq1}), we find that
\begin{equation}
24\lambda = \sum^\infty_{k=0} \lambda^{k+1}
\left[ 16\lambda^4 \frac{d^4}{d\xi^4} S_k(0) + 80 \lambda^3 \frac{d^3}{d\xi^3} S_k(0)
+ 140 \lambda^2 \frac{d^2}{d\xi^2} S_k(0)
+ 100 \lambda \frac{d}{d\xi} S_k(0) + 24 S_k(0)\right]~.
\label{eq13}
\end{equation}
(We note that $L = 0$ if $\phi = \mu$.) It follows from Eq.\  (\ref{eq13}) that for the CW renormalization scheme
\begin{gather}
S_0(0) = T_{00}=1
\label{eq14}
\\
0=100 S_0^\prime (0) + 24 S_1 (0) = 50b_2 T_{00}+24 T_{10} \Longrightarrow T_{10}=-\frac{25}{12}b_2
\label{eq15}
\\
0=140 S_0^{\prime\prime}(0)+100S_1^\prime(0)+24S_2(0)= 12 T_{20}+50b_2 T_{10}+\left(35b_2^2+25b_3+100g_2\right) T_{00} 
\label{boundary_eq2}
\\
\begin{split}
0=&80S_0^{\prime\prime\prime}(0)+140 S_1^{\prime\prime}(0)+100S_2^\prime(0)+24S_3(0)
\\
=&\left(60b_2^3+175b_2b_3+50b_4+610 b_2g_2+200g_3 \right)T_{00}
\\
&+\left(210b_2^3+100b_3-48\frac{b4}{b_2}+200g_2-48\frac{g_3}{b_2}+\frac{48}{b_2}\left[b_4+g_3\right]\right)T_{10}
+150b_2T_{20}+24 T_{30}
\label{boundary_eq4}
\end{split}
\end{gather}
{\it etc.}; in general by having Eq.\  (\ref{eq13}) satisfied at each order in $\lambda$ we end up with
\begin{equation}
16 \frac{d^4}{d\xi^4}S_k (0) + 80 \frac{d^3}{d\xi^3}S_{k+1}(0) + 140 \frac{d^2}{d\xi^2}S_{k+2}(0)
+ 100 \frac{d}{d\xi}S_{k+3} (0) + 24 S_{k+4} (0) = 0~(k=0,1,2 \ldots)~.
\label{eq16}
\end{equation}
Consequently, the ($n+1$ loop) boundary condition $S_n(0)=T_{n0}$ is  determined  iteratively by lower-order results via Eq.\  (\ref{eq16}); that is, once $S_k(\xi)\, \ldots \,S_{k+3}(\xi)$ are known,  $S_{k+4}(0)=T_{k+4 \,0}$ is fixed by Eq.\  (\ref{eq16}). 
Hence $V$ is determined entirely by the renormalization group functions $\beta(\lambda)$, $\gamma(\lambda)$ when employing the CW renormalization condition of Eq.~(\ref{eq1}). It is not apparent how $T_{n0}$ can be determined in any other scheme except by relating that scheme to the CW scheme.

However, the renormalization group functions are generally computed in the MS renormalization scheme; being given up to five loop order in Ref.\ \cite{9}. To relate the renormalization group functions in these two schemes, we note, following Ref.\ \cite{7}, that in the MS scheme, the form of the effective potential is much the same as that of Eq.\  (\ref{eq2}),
\begin{equation}
V(\lambda, \phi, \tilde{\mu}) = \sum_{n=0}^\infty \sum_{m=0}^n \lambda^{n+1} \tilde{T}_{nm}\tilde{L}^m\phi^4
\label{eq17}
\end{equation}
where now
\begin{equation}
\tilde{L} = \log\left(\frac{\lambda\phi^2}{\tilde{\mu}^2}\right)~.
\label{eq18}
\end{equation}
Upon comparing Eqs.\ (\ref{eq3}) and (\ref{eq18}), the mass scale $\tilde{\mu}^2$ in the MS scheme can be related to the mass scale $\mu^2$ in the scheme of Eq.\  (\ref{eq1}) by
\begin{equation}
\tilde{\mu} = \lambda^\frac{1}{2}  \mu
\label{eq19}
\end{equation}
 Consequently,
$\frac{d\mu}{d\tilde{\mu}} = \lambda^{-\frac{1}{2}} - \frac{\lambda^{-3/2}}{2} \tilde{\beta}(\lambda)$ where
$\tilde{\beta}(\lambda) = \tilde{\mu} \frac{d\lambda}{d\tilde{\mu}}$, $\phi\tilde{\gamma}(\lambda) = \tilde{\mu} \frac{d\phi}{d\tilde{\mu}}$
and thus \cite{7}
\begin{gather}
\beta(\lambda) = \frac{\tilde{\beta}(\lambda)}{1 - \frac{\tilde{\beta}(\lambda)}{2\lambda}}
\label{eq20}
\\
\gamma(\lambda) = \frac{\tilde{\gamma}(\lambda)}{1 - \frac{\tilde{\beta}(\lambda)}{2\lambda}}~
\label{eq21}
\end{gather}
relates the renormalization group function in the two schemes. Knowing $\tilde{\beta}(\lambda)$ and $\tilde{\gamma}(\lambda)$ in the MS renormalization scheme thus determines $\beta(\lambda)$ and $\gamma(\lambda)$ in the CW renormalization scheme 
[Eq.~(\ref{eq1})] and hence $V$ can be determined entirely from $\tilde{\beta}(\lambda)$ and $\tilde{\gamma}(\lambda)$.
In particular, if $\tilde{\beta}(\lambda) = \tilde{b}_2 \lambda^2 + \tilde{b}_3\lambda^3 + \ldots$,
$\tilde{\gamma}(\lambda) = \tilde{g}_1 \lambda + \tilde{g}_2\lambda^2 + \ldots$, then 
Eqs.~(\ref{eq20}) and (\ref{eq21}) can be expanded  to convert the five-loop MS-scheme renormalization group functions of \cite{9}. The explicit results to three-loop order (the first order at which the conversion of the anomalous dimension is non-trivial) are  
{\allowdisplaybreaks
\begin{gather}
b_2 = \tilde{b}_2
\label{b2_trans}
\\
b_3 = \tilde{b}_3 + \frac{1}{2} \tilde{b}_2^2
\label{b3_trans}
\\
b_4 = \tilde{b}_4 + \tilde{b}_2\tilde{b}_3 + \frac{1}{4} \tilde{b}_2^3
\\
g_1=\tilde g_1=0\\
g_2 = \tilde{g}_2
\label{g2_trans}
\\ 
g_3 = \tilde{g}_3 + \frac{1}{2}\tilde{b}_2 \tilde{g}_2
\end{gather}
}
The results up to five loop order (or indeed any desired order) are easily obtained.

The MS-scheme RG coefficients  for the $O(N)$ version of the $\lambda\phi^4$ model are known to five-loop order \cite{9}; to establish our conventions their values to three-loop order are
{\allowdisplaybreaks
\begin{gather}
\tilde{b}_2= \frac{N+8}{2\pi^2}
\label{b2_ms}
\\
\tilde{b}_3 = -\frac{3(3N+14)}{4\pi^4}
\label{b3_ms}
\\
\tilde{b}_4 = \frac{33N^2 + 922N + 2960 + 96(5N+22)\zeta(3)}{64\pi^6}
\\
\tilde{g}_1=0
\\
\tilde{g}_2 = -\frac{N+2}{16\pi^4}
\label{g2_ms}
\\
\tilde{g}_3 = \frac{(N+2)(N+8)}{128\pi^6}\, .
\end{gather}
}

The sum of all leading-logarithm (LL) and next-to-leading-logarithm (NLL) contributions to $V$ is given by
\begin{equation}
V_{LL+NLL} = \lambda\left[S_0(\lambda L) + \lambda S_1(\lambda L)\right]\phi^4 
\end{equation}
where $S_0$ and $S_1$ are completely determined by Eqs.~(\ref{S0_sol}), (\ref{S1_sol}), (\ref{eq14}), (\ref{eq15}), 
(\ref{b2_trans}), (\ref{b3_trans}), (\ref{g2_trans}), (\ref{b2_ms}), (\ref{b3_ms}), (\ref{g2_ms}). 
We can then recover the complete two-loop CW-scheme result for $V$ computed explicitly in Ref.\ \cite{3} upon expanding $S_0$ and $S_1$ to terms quadratic in $L=\log\frac{\phi^2}{\mu^2}$  and  calculating $T_{20}$ via (\ref{boundary_eq2}).  
It should be noted that the two-loop CW-scheme result \cite{3} does not satisfy the RG equation with MS coefficients; if one applies solves the RG equation with MS-scheme RG functions it would disagree with the explicit two-loop calculation. 

As noted earlier, $S_4$ requires knowledge of the renormalization-group functions to five-loop order, and hence $S_4$ is the highest-order term in the expansion (\ref{eq7}) that can be determined with current knowledge of the renormalization group functions in the massless $O(N)$ theory \cite{9}.
The solutions $S_0$, $S_1$, $S_2$, $S_3$ and $S_4$ contain the boundary-condition coefficients $T_{n0}$ for $n\le 4$.  These five coefficients are determined by the set of five equations (\ref{eq14})--(\ref{boundary_eq4}) and  (\ref{eq16}) with $k=0$. Once these coefficients are determined, the $N^4LL$ expression for the effective potential
\begin{equation}
V_{N^4LL} = \sum^4_{n=0} \lambda^{n+1} S_n (\lambda L)\,\phi^4
\end{equation}
is expanded up to fifth order in $L$  to obtain the five-loop perturbative coefficients $T_{nm}$ for $0\leq n\leq 5$ and $ 0< m \leq 5$.  The remaining five-loop coefficient $T_{50}=S_5(0)$ is determined via (\ref{eq16}) with $k=1$.
We present our determination of the explicit values for the perturbative coefficients to three-loop order
(one order higher than the CW scheme calculation of Ref.\ \cite{3})
\begin{gather}
T_{00}=1 \label{tree_T}\\
T_{10}=-\frac{25(N+8)}{24\pi^2}~,~T_{11}=\frac{N+8}{4\pi^2}\\
T_{20}=\frac{5\left( 17N^2+347N+1418\right)}{72\pi^4}~,~
T_{21}=-\frac{11N^2+206N+836}{24\pi^4}~,~
T_{22}=\frac{\left(N+8\right)^2}{16\pi^4} 
\label{two_loop_T}
\\
T_{30}=-\frac{5\left[
784 N^3+26305 N^2+\left[251338+7200\zeta(3) \right] N+694032+31680\zeta(3)
\right]}{2304\pi^6}\\
T_{31}=\frac{296N^3+9425N^2+\left[87242+1440\zeta(3)\right]N+239376+6336\zeta(3)}{384\pi^6}
\\
T_{32}=-\frac{(N+8)\left(10N^2+209N+858\right)}{64\pi^6}
~,~T_{33}=\frac{(N+8)^3}{64\pi^6}\, .
\end{gather}
The results (\ref{tree_T})--(\ref{two_loop_T}) are in agreement with the explicit two-loop calculation \cite{3}.
As mentioned earlier, the corresponding expression $S_2(\xi)$ used to obtain  these coefficients  is given in Appendix A. Although the analytic expressions for the remaining coefficients to five-loop order are too lengthy to be presented, we give their numerical values for $N=1$ (simple scalar field theory) and $N=4$ (the scalar-field theory projection of the Standard Model) in Tables \ref{four_coeff_tab} and \ref{five_coeff_tab}.   

\begin{table}[hbt]
\centering
\begin{tabular}{||c|c|c|c|c|c||}
 \hline\hline
 & $T_{40}$ & $T_{41}$ & $T_{42}$ & $T_{43}$ & $T_{44}$ \\ \hline\hline
$N=1$ &  $5.218$ & $-2.277$ & $0.4457$ & $-4.865\times 10^{-2}$ & $2.701\times 10^{-3}$
\\
\hline
$N=4$ & $14.59$ & $-6.477$ & $1.304$ & $-0.1475$ & $8.537\times 10^{-3}$
\\
\hline \hline
\end{tabular}
\caption{Four-loop perturbative coefficients for $N=1$ and $N=4$.} 
\label{four_coeff_tab}
\end{table}
 
\begin{table}[hbt]
\centering
\begin{tabular}{||c|c|c|c|c|c|c||}
 \hline\hline
 & $T_{50}$ & $T_{51}$ & $T_{52}$ & $T_{53}$ & $T_{54}$ & $T_{55}$ \\ \hline\hline
$N=1$ & $-14.06$  & $6.407$  & $-1.362$  & $0.1744$ & $-1.404\times 10^{-2}$ & $6.158\times 10^{-4}$
\\
\hline
$N=4$ & $-50.09$ & $23.20$ & $-5.065$  & $0.6709$ & $-5.631\times 10^{-2}$  & $2.595\times 10^{-3} $
\\
\hline \hline
\end{tabular}
\caption{Five-loop perturbative coefficients for $N=1$ and $N=4$.} 
\label{five_coeff_tab}
\end{table}

In the next Section, we examine how the methods developed for massless scalar field theory  can be extended to massless scalar electrodynamics, a theory with multiple couplings. 

\section{Massless Scalar Electrodynamics}\label{sqed_sec}
Massless scalar quantum electrodynamics (MSQED) has the Lagrangian
\begin{equation}
L = \frac{1}{2} \left[ \left(\partial_\mu + i e A_\mu \right) \phi^* \right] \left[ \left( \partial^\mu - i e A^\mu \right) \phi \right] - \frac{1}{4} \left( \partial_\mu A_\nu - \partial_\nu A_\mu \right)^2 - \frac{g}{4 !} \left( \phi^* \phi \right)^2
\label{sqed_eq1}  
\end{equation}
The effective potential $V(\phi)$  in this model can be computed perturbatively in a variety of ways \cite{1,2,3,4,salam}.  The effective action for MSQED has the expansion
\begin{equation}
 \Gamma=\int d^4x\, \left[
-V(\phi)-\frac{1}{4}H(\phi)F_{\mu\nu}F^{\mu\nu}
+\frac{1}{2}Z(\phi)\left(\partial_\mu + i e A_\mu \right) \phi^*\left(\partial_\mu - i e A_\mu \right) \phi
+\ldots
\right]\, .
\end{equation}
In the CW renormalization condition \cite{1}
\begin{gather}
\left.\frac{d^4V(\phi)}{d \phi^4} \right| _{\phi=\mu} = g \, ,
\label{sqed_eq2}
\\
\left. H(\phi) \right| _{\phi=\mu}=1=\left. Z(\phi) \right| _{\phi=\mu} \, ,
\end{gather}
one finds a perturbative expansion for $V(\phi)$ of the form
\begin{gather}
V(g,\alpha,\phi, mu)= \left(\sum^\infty_{n=1} \sum^{n+k}_{r=0} \sum^\infty_{k=0} T_{n+k-r, r, k} g^{n+k-r} \alpha^r L^k \right) \phi^4 \, ,
\label{sqed_eq3}
\\
L=\log{\left( \frac{\phi^2}{\mu^2} \right)}\, ,
\label{sqed_cw_logdef}
\end{gather}
where $\alpha = e^2$, and $\mu^2$ is the RG scale appearing in Eq.~(\ref{sqed_eq2}).  The effective potential  satisfies the RG equation
\begin{equation}
\mu \frac{dV}{d \mu} = 0 = \left( \mu \frac{\partial}{\partial \mu} + \beta^g \frac{\partial}{\partial g} + \beta^\alpha \frac{\partial}{\partial \alpha} +\gamma\phi\frac{\partial}{\partial\phi}\right) V \left( g, \alpha,  \phi ,\mu \right)\equiv D V~.
\label{sqed_eq4} 
\end{equation}
Here, $\beta^g, \beta^\alpha $ and $\gamma$ are the renormalization group functions
\begin{gather}
\beta^g (g, \alpha) = \mu \frac{d g}{d \mu} = \sum^\infty_{n=2} \beta^g_n 
\quad ,\quad \beta^g_n = \sum^n_{r=0} b^g_{n-r, r} g^{n-r} \alpha^r  
\label{sqed_eq5}
\\
\beta^\alpha (g, \alpha) = \mu \frac{d \alpha}{d \mu} = \sum^\infty_{n=2} \beta^\alpha_n 
\quad ,\quad  \beta^\alpha_n = \sum^n_{r=0} b^\alpha_{n-r, r} g^{n-r} \alpha^r  
\label{sqed_eq6}
\\
\gamma (g, \alpha) = \frac{\mu}{\phi} \frac{d \phi}{d \mu} = \sum^\infty_{n=1} \gamma_n 
\quad , \quad  \gamma_n = \sum^n_{r=0} \gamma_{n-r, r} g^{n-r} \alpha^r 
\label{sqed_eq7}
\end{gather}
In the previous Section it was shown that the effective potential in an $O(N)$-symmetric massless $\lambda\phi^4$ theory is uniquely determined by the RG equation in the CW scheme.  We now extend this analysis to deal with the situation occurring in MSQED where two couplings $g$ and $\alpha$ occur.  
As shown below, there are two crucial distinctions between the single- and multiple-coupling situations.  First, the coupled ordinary differential equations (\ref{neweq11}) get replaced by coupled partial differential equations.  Secondly, the conversion of the RG functions from the MS scheme to the CW scheme requires use of multi-scale RG methods \cite{multi_scale_einhorn,multi_scale_ford}.

We now proceed to shown how together Eqs.\ (\ref{sqed_eq2}) and (\ref{sqed_eq4}) again determine $V$ without the calculation of additional Feynman diagrams.  We first define
\begin{equation}
p^k_n (g, \alpha) = \sum^n_{r=0} T_{n-r, r, k} g^{n-r} \alpha^r \; (n \geq k + 1) 
\label{sqed_eq8} 
\end{equation}
As a result, Eq.\ (\ref{sqed_eq3}) can be written
\begin{equation}
V (g,\alpha,\phi,\mu) = \sum^\infty_{n=1} \sum^{n-1}_{k=0} p^k_n ( g, \alpha) L^k \phi^4 
\label{sqed_eq9}\, , 
\end{equation}
with the contributions
\begin{gather}
V_{LL} = \sum^\infty_{k=0} p^k_{k+1} L^k \phi^4 \, ,
\label{sqed_eq10} 
\\
V_{NLL} = \sum^\infty_{k=0} p^k_{k+2} L^k \phi^4 \, \ldots
\label{sqed_eq11} 
\\
V_{N^pLL} = \sum^\infty_{k=0} p^k_{k+p+1} L^k \phi^4 
\label{sqed_eq12} 
\end{gather}
giving the leading-log, next-to-leading-log {\it etc.\ }corrections to $V$.

First, we find that substitution of (\ref{sqed_eq9}) into (\ref{sqed_eq2}) results in the condition
\begin{equation}
24p^0_n + 100 p^1_n + 280p^2_n+480p^3_n+384p^4_n= g 
\label{sqed_eq13} 
\end{equation}
upon noting that if $\phi = \mu$, then $L = 0$.  Recalling that $p^k_n=0$ if $n < k + 1$, we see that if $n=1$, Eq.\ (\ref{sqed_eq13}) leads to
\begin{equation}
p^0_1 = \frac{g}{24}\,, 
\label{sqed_eq14} 
\end{equation}
the tree level result.

We now substitute Eqs.\ (\ref{sqed_eq5})--(\ref{sqed_eq7}), (\ref{sqed_eq9}) into Eq.\ (\ref{sqed_eq4}) to obtain
\begin{equation}
\sum^\infty_{n=1} \sum^{n-1}_{k=0} \left[ -2 k p^k_n L^{k-1} + \sum^\infty_{m=2} \left( \beta^g_m \frac{\partial p^k_n}{\partial g} + \beta^\alpha_m \frac{\partial p^k_n}{\partial \alpha} \right)  L^k 
+ \sum^\infty_{m=1} \left(4 \gamma_m p^k_n L^k + 2 k \gamma_m p^k_n L^{k-1} \right) \right] \phi^4 = 0 
\label{sqed_eq15}
\end{equation}
Since $p^k_n, \beta^g_n, \beta^\alpha_n$ and $\gamma_n$ are all polynomials of degree $n$ in $g$ and $\alpha$, we can obtain coupled first-order partial differential equations that express each of the $p^k_n$ in terms of $\beta^g_n$, $\beta^\alpha_n$ and $\gamma_n$.  This is done by requiring that Eq.\ (\ref{sqed_eq15}) be satisfied order-by-order in $L^k$ and in $\sum^n_{r=0} c_r g^{n-r} \alpha^r$.  These equations are first order partial differential equations whose boundary conditions are provided by Eq.\ (\ref{sqed_eq13}). 

For example, at second order in the couplings and order zero in $L$, Eq.\ (\ref{sqed_eq15}) leads to
\begin{equation}
-2p^1_2 + \beta^g_2 \frac{\partial p^0_1}{\partial g} + \beta^\alpha_2 \frac{\partial p^0_1}{\partial\alpha} + 4 \gamma_1 p^0_1 = 0 \, .
\label{sqed_eq16}
\end{equation}
This fixes $p^1_2$, since $p^0_1$ is given by Eq.\ (\ref{sqed_eq14}).  Now at third order in the couplings and order $L$, we see from Eq.\ (\ref{sqed_eq15}) that
\begin{equation}
-4 p^2_3 + \beta^g_2 \frac{\partial p^1_2}{\partial g} + \beta^\alpha_2 \frac{\partial p^1_2}{\partial\alpha} + 4 \gamma_1 p^1_2 = 0 \, .
\label{sqed_eq17}
\end{equation}
Since $p^1_2$ has been determined, Eq.\ (\ref{sqed_eq17}) serves to fix $p^2_3$. 
In general, at order $(n+2)$ in the couplings, and order $n$ in $L$, we find that
\begin{equation}
p^{n+1}_{n+2} = \frac{1}{2(n+1)} \left[ \beta^g_2 \frac{\partial}{\partial g} + \beta^\alpha_2 \frac{\partial}{\partial \alpha} + 4 \gamma_1 \right] p^n_{n+1}\, , 
\label{sqed_eq18}
\end{equation}
so that all of the contributions to $V_{LL}$ in Eq.\ (\ref{sqed_eq10}) can be calculated provided we use the expression for $p^0_1$ given by Eq.\ (\ref{sqed_eq14}).

Since we now know $p^1_2$, we can set $n=2$ in Eq.\ (\ref{sqed_eq13}), leading to
\begin{equation}
 p^0_2 = - \frac{25}{6} p^1_2 \, .
\label{sqed_eq19}
\end{equation}
This serves to start the sequence $p^0_2, p^1_3 \dots$ {\it etc.} Upon looking at terms in Eq.\ (\ref{sqed_eq15}) that are of order $n+3$ in the couplings and $n$ in $L$ we have the recursion relation
\begin{equation}
\left(p^n_{n+2} - \gamma_1 p^n_{n+1} \right) = \frac{1}{2n} \left[ \left(\beta^g_2 \frac{\partial}{\partial g} + \beta^\alpha_2 \frac{\partial}{\partial \alpha} + 4 \gamma_1 \right) p^{n-1}_{n+1} + \left( \beta^g_3 \frac{\partial}{\partial g} + \beta^\alpha_3 \frac{\partial}{\partial \alpha} + 4 \gamma_2 \right) p^{n-1}_n \right] \, ,
\label{sqed_eq20} 
\end{equation}
which finally fixes all contributions to $V_{NLL}$ in Eq.\ (\ref{sqed_eq11}) in terms of $\beta^g_2$, 
$\beta^g_3$, $\beta^\alpha_2$, $\beta^\alpha_3$, $\gamma_1$ and $\gamma_2$. 

It is evident that $V_{N^PLL}$ involves the polynomials $p^n_{n+p+1}$.  From Eq.\ (\ref{sqed_eq13}), $p^0_{p+1}$ can be found once $p^1_{p+1} \dots p^4_{p+1}$ have been computed in the course of determining $V_{N^{p-1}LL} \dots V_{LL}$.  Having fixed $p^0_{p+1}$ in this way, all subsequent contributions to $V_{N^PLL}$ are determined in terms of the polynomials $p^n_{n+p} \dots p^n_{n+1}$ as well as $\beta^g_2 \dots \beta^g_{p+1}, \beta^\alpha_2 \dots \beta^\alpha_{p+1}, \gamma_1 \dots \gamma_p$, by considering those terms in Eq.\ (\ref{sqed_eq15}) that are of order $n+p+2$ in the coupling, and order $n$ in $L$.  
The effective potential for massless scalar QED is therefore completely determined by the RG functions in the CW renormalization scheme.
In Appendix B, the sums appearing in Eqs.\ (\ref{sqed_eq10}), (\ref{sqed_eq11}) for $V_{LL}$
and $V_{NLL}$ are evaluated in closed form using  a variant of the method of characteristics.

As in the case of $O(N)$-symmetric massless $\lambda \phi^4$ theory, it is necessary to convert the RG functions from the MS scheme to the CW scheme.  However, because there are two logarithms 
$\log\left(g\phi^2/\tilde\mu^2\right)$ and $\log\left(\alpha\phi^2/\tilde\mu^2\right)$ 
appearing in the MS-scheme perturbative expansion, 
a simple rescaling  of the renormalization scale $\tilde\mu$ as in (\ref{eq19}) cannot convert these two logarithms 
into the single logarithm (\ref{sqed_cw_logdef}).  

The presence of multiple incompatible logarithmic scales is known to
cause difficulties when attempting to solve the RG equation in other applications. To circumvent these problems, the concept of multiple renormalization scales, one scale for each appearance of the traditional MS RG scale in the Lagrangian, was first considered in \cite{multi_scale_einhorn}.  This method was refined in \cite{multi_scale_ford} by associating a renormalization scale with each kinetic term in the Lagrangian, which in the case of massless scalar QED, will introduce two renormalization scales ($\kappa_g$ and $\kappa_\alpha$) resulting in a MS-scheme perturbation series for the effective potential containing two logarithms:
\begin{equation}
 L_g=\log{\left(\frac{g \phi^2}{\kappa_g^2}\right)} 
~,~
L_\alpha= \log{\left(\frac{\alpha \phi^2}{\kappa_\alpha^2}\right)}\, .
\label{multi_log_def}
\end{equation}
With multiple renormalization scales, there will exist MS-scheme RG equations and RG functions associated with each scale \cite{multi_scale_ford}
\begin{gather}
\kappa_g \frac{dV}{d \kappa_g} = 0 = \left( \kappa_g \frac{\partial}{\partial \kappa_g} + \tilde\beta^g_g \frac{\partial}{\partial g} + \tilde\beta^\alpha_g \frac{\partial}{\partial \alpha} +\tilde\gamma_g\phi\frac{\partial}{\partial\phi}\right) V \equiv D_1 V~,
\\
\kappa_\alpha \frac{dV}{d \kappa_\alpha} = 0 = \left( \kappa_\alpha \frac{\partial}{\partial \kappa_\alpha} + \tilde\beta^g_\alpha \frac{\partial}{\partial g} + \tilde\beta^\alpha_\alpha \frac{\partial}{\partial \alpha} +\tilde\gamma_\alpha\phi\frac{\partial}{\partial\phi}\right) V \equiv D_2 V~,
\end{gather}
where
\begin{gather}
\tilde\beta^g_g=\kappa_g\frac{\partial g}{\partial \kappa_g}~,~ \tilde\beta^g_\alpha=\kappa_\alpha\frac{\partial g}{\partial \kappa_\alpha}
\label{multi_RG_beta_g}
\\
\tilde\beta^\alpha_g=\kappa_g\frac{\partial \alpha}{\partial \kappa_g}~,~ \tilde\beta^\alpha_\alpha=\kappa_\alpha\frac{\partial \alpha}{\partial \kappa_\alpha}
\label{multi_RG_beta_alpha}
\\
\tilde\gamma_g\phi=\kappa_g\frac{\partial\phi}{\partial\kappa_g}~,~\tilde\gamma_\alpha\phi=\kappa_\alpha\frac{\partial\phi}{\partial\kappa_\alpha}\, .
\label{multi_RG_gamma}
\end{gather}
As outlined in Ref.\ \cite{multi_scale_ford}, these multi-scale MS-scheme RG functions can be obtained from the $1/\epsilon$ poles in the (multi-scale) renormalization constants. These multi-scale RG functions can also be determined by reconstructing the effective potential in the MS scheme from the MS RG functions and the logarithm-free parts of $V$; at this stage the renormalization scale $\tilde\mu$ can be split into $\kappa_g$ and $\kappa_\alpha$ allowing for a determination of Eqs.~(\ref{multi_RG_beta_g})--(\ref{multi_RG_gamma}) through the requirement that $V$ be independent of
both $\kappa_g$ and $\kappa_\alpha$ along the lines of Ref.~\cite{more_gerry}.   Furthermore,
in the limit when the two scales coincide ($\kappa_g=\kappa_\alpha=\tilde\mu$), the multi-scale MS RG functions are related to the single-scale MS RG functions $\tilde\beta^g$, $\tilde\beta^\alpha$ and $\tilde\gamma$ via \cite{multi_scale_ford}
\begin{equation}
 \tilde\beta^g=\tilde\beta^g_g+\tilde\beta^g_\alpha~,~\tilde\beta^\alpha=\tilde\beta^\alpha_g+\tilde\beta^\alpha_\alpha
~,~\tilde\gamma=\tilde\gamma_g+\tilde\gamma_\alpha\, .
\label{multi_scale_bc}
\end{equation}
The RG functions (\ref{multi_RG_beta_g})--(\ref{multi_RG_gamma}) must also be consistent with the integrability condition 
$\left[ D_1,D_2\right]V=0$; this constraint combined with the boundary condition (\ref{multi_scale_bc}) may also be used to determine the multi-scale RG functions \cite{multi_scale_ford}. 

It is now evident that the rescalings
\begin{equation}
 \kappa_g=\sqrt{g}\mu~,~\kappa_\alpha=\sqrt{\alpha}\mu \, ,
\label{sqed_rescale}
\end{equation}
will convert the MS-scheme multi-scale logarithms (\ref{multi_log_def}) into the CW-scheme logarithm (\ref{sqed_cw_logdef}), thereby enabling scheme conversion.  The RG functions in the CW scheme can then be obtained from (\ref{sqed_rescale}) combined with
\begin{gather}
\beta^g=\mu\frac{dg}{d\mu}=\mu\frac{d \kappa_g}{d \mu}\frac{\partial g}{\partial \kappa_g}
+ \mu\frac{d \kappa_\alpha}{d \mu}\frac{\partial g}{\partial \kappa_\alpha}
\\
\beta^\alpha=\mu\frac{d\alpha}{d\mu}=\mu\frac{d \kappa_g}{d \mu}\frac{\partial \alpha}{\partial \kappa_g}
+ \mu\frac{d \kappa_\alpha}{d \mu}\frac{\partial \alpha}{\partial \kappa_\alpha}
\\
\gamma\phi=\mu\frac{d\phi}{d\mu}=\mu \frac{d\kappa_g}{d \mu}\frac{\partial \phi}{\partial \kappa_g}
+ \mu \frac{d\kappa_\alpha}{d \mu}\frac{\partial \phi}{\partial \kappa_\alpha}\, ,
\end{gather}
to obtain
\begin{gather}
 \beta^g=\tilde\beta^g_g\left[ 1+\frac{\beta^g}{2g}\right]
+\tilde\beta^g_\alpha\left[ 1+\frac{\beta^\alpha}{2\alpha}\right]
\label{sqed_convert1}
\\
\beta^\alpha=\tilde\beta^\alpha_g\left[ 1+\frac{\beta^g}{2g}\right]
+\tilde\beta^\alpha_\alpha\left[ 1+\frac{\beta^\alpha}{2\alpha}\right]
\label{sqed_convert2}
\\
\gamma=\tilde\gamma_g \left[ 1+\frac{\beta^g}{2g}\right]+ \tilde\gamma_\alpha \left[ 1+\frac{\beta^\alpha}{2\alpha}\right]\, .
\label{sqed_convert3}
\end{gather}
The above equations can be solved perturbatively for the coefficients of the CW-scheme RG functions in terms of the multi-scale MS-scheme RG functions.\footnote{It can be verified that the scalar field theory scheme conversion results (\ref{eq20})--(\ref{eq21}) are obtained from the $\alpha\to 0$ limit of the MSQED results (\ref{sqed_convert1})--(\ref{sqed_convert3}) .  In this limit, $\tilde\beta_g^g\to \tilde\beta$, 
$\tilde\gamma_g\to \tilde\gamma$, and all other multi-scale MS RG functions become zero.  Inversion of the 
resulting expressions  $\beta^g=\tilde\beta\left[1+\beta^g/(2g)\right]$ and 
$\gamma=\tilde\gamma \left[1+\beta^g/(2g)\right]$ lead to $\beta^g=\tilde\beta/\left[1-\tilde\beta/(2g)\right]$
and $\gamma=\tilde\gamma/\left[1-\tilde\beta/(2g)\right]$ consistent with Eqs.~(\ref{eq20})--(\ref{eq21}).}
As expected, to lowest order one finds that the CW-scheme and MS-scheme RG coefficients coincide so that the effects of scheme conversion enter at two-loop level.

\section{Conclusions}
In summary, we have
developed iterative techniques that uniquely determine, in terms of MS RG functions, leading-logarithm and subsequent-to-leading-logarithm expansions 
 of the effective potential in the CW scheme for massless $\lambda \phi^4$ scalar field theory with a 
global $O(N)$ symmetry.  In these techniques, the $N^pLL$ expression  is governed by a coupled set of first-order ordinary differential equations containing the $p+1$ loop RG coefficients, and the boundary conditions for this system are determined by the CW renormalization condition.  In this approach, it 
is essential to convert the RG functions from the MS scheme (in which they are known to five-loop order in $O(N)$-symmetric massless scalar field theory) to the CW-scheme.

The methods developed for the scalar field theory one-coupling case have been extended to massless scalar QED.  The presence of two couplings does not change the essential features of the analysis; instead of coupled ordinary differential equations the $N^pLL$ expansions are determined by systems of first-order partial differential equations resulting from the RG equation and algebraic equations arising from the CW renormalization condition. Similarly,  conversion of the RG functions from the MS scheme to the CW scheme in massless scalar QED is also more elaborate, and requires the use of multi-scale renormalization group methods.   
Although  multi-scale RG techniques are not widely known, the necessary multi-scale RG functions can either be calculated directly by introducing a renormalization scale for each kinetic term (and hence propagator) in the theory and exploiting the usual relation between the RG functions and the $1/\epsilon$ terms in the renormalization constants, or they may be reconstructed from the single-scale MS-scheme RG functions in conjunction with integrability conditions related to the commutator of the RG operator associated with each renormalization scale 
\cite{multi_scale_einhorn,multi_scale_ford}.

We would like to extend our methods to computing the effective potential when the mass of the field $\phi$ is non-zero. In particular, 
our analysis may allow us to correct the two-loop renormalization-group analysis of the Standard Model effective potential   
 Ref.~\cite{10} which is in disagreement with the explicit two-loop calculation \cite{11}.
It would also be interesting to see if the effective potential in the MS renormalization scheme could be determined uniquely by the renormalization group functions.

\section*{Acknowledgements}
We all especially want to express our indebtedness to the late Dr.~Victor Elias, whose insights led directly to the results presented here.  D.G.C.~McKeon and F.~Chishtie would like to thank the University of Saskatchewan for its hospitality while this work was being done. NSERC provided financial support.

\section*{Appendix A: Massless $\lambda \phi^4$ Theory}
The differential equation (\ref{neweq11}) establishes
a recursive relation for the 
$S_n(\xi)$   
resulting in solutions of the form
\begin{equation}
S_n(\xi)=\frac{1}{b_2}\sum_{i=1}^{n+1}\sum_{j=0}^{i-1}\sigma_{i,j}^{{n}}\frac{L^j}{w^i}
\label{form}
\end{equation}
where $w = 1 - \frac{b_2}{2} \xi$, $\xi=\lambda L$, and $L\equiv \log(w)$.   

In terms of this notation, the solution for $S_0$ is 
\begin{equation}
S_0(\xi)=\frac{\sigma_{1,0}^{{0}}}{b_2w} ~,~\sigma_{1,0}^{{0}}=b_2T_{00}=b_2~,
\end{equation}
and the solution for $S_1$ is 
\begin{equation}
S_1(\xi)=\frac{1}{b_2}\left(\frac{\sigma_{1,0}^{{1}}}{w}+\frac{\sigma_{2,0}^{{1}}}{w^2}+\frac{\sigma_{2,1}^{{1}}L}{w^2}
\right)~,
\end{equation}
where
\begin{equation}
\sigma_{1,0}^{{1}}=-4 g_2 T_{00}~,~\sigma_{2,0}^{{1}}=
b_2T_{10}+4g_2T_{00}~,~ \sigma_{2,1}^{{1}} = -b_3T_{00}  ~.
\end{equation}

For the higher-order $S_n$, recursive expressions for $\sigma^n_{i,j}$ provide the most compact form.  For example, 
the solution for $S_2$ is:
\begin{equation}
S_2(\xi)=\frac{1}{b_2}\left(\frac{\sigma_{1,0}^{{2}}}{w}+\frac{\sigma_{2,0}^{{2}}}{w^2}+\frac{\sigma_{2,1}^{{2}}L}{w^2}
+\frac{\sigma_{3,0}^{{2}}}{w^3}+\frac{\sigma_{3,1}^{{2}}L}{w^3}+\frac{\sigma_{3,2}^{{2}}L^2}{w^3}\right)
\end{equation}
where
\begin{gather}
\sigma_{1,0}^{{2}}=-\frac{1}{2}(b_3+4g_2)\sigma_{1,0}^{{1}}-2T_{00}g_3~, \\
\sigma_{2,0}^{{2}}=-\left\{b_3\sigma_{1,0}^{{1}}+4g_2\sigma_{2,0}^{{1}}+(b_3-4g_2)\sigma_{2,1}^{{1}}+(b_4+b_2g_2)T_{00}\right\}~,\\
\sigma_{2,1}^{{2}} = -4g_2\sigma_{2,1}^{{1}}~, \\
\sigma_{3,0}^{{2}}=\frac{1}{2}\left\{(3b_3+4g_2)\sigma_{1,0}^{{1}}+8g_2\sigma_{2,0}^{{1}}+(2b_3-8g_2)\sigma_{2,1}^{{1}}+
(2b_4+4g_3+2b_2g_2)T_{00}+2b_2T_{20}\right\}~,\\
\sigma_{3,1}^{{2}}=-b_3(2\sigma_{2,0}^{{1}}-\sigma_{2,1}^{{1}})~,
\\
\sigma_{3,2}^{{2}}=-b_3\sigma_{2,1}^{{1}}~.
\end{gather}

The differential equation (\ref{neweq11}) combined with the form of the solution (\ref{form})
can be used to
obtain a set of recursion relations for the coefficients
$\sigma_{i,j}^{{n}}$. One finds that this procedure yields
\begin{equation}
\begin{split}
0=&b_2(j+1)\sigma_{i,j+1}^{{n}}+\{(n-i+1)b_2+4g_1\}\sigma_{i,j}^{{n}}
+\sum_{m=0}^{n-1}[b_{n+2-m}(j+1)\sigma_{i,j+1}^{{m}}+(i-1)(b_2g_{n-m}+b_{n+2-m})\sigma_{i-1,j}^{{m}} \\
&-(j+1)(b_2g_{n-m}+b_{n+2-m})\sigma_{i-1,j+1}^{{m}}+(4g_{n+1-m}+(m-i+1)b_{n+2-m})\sigma_{i,j}^{{m}}]
\end{split}
\label{recur}
\end{equation}
We note that $\sigma_{i,j}^{{n}}=0$ if $i>n+1, ~j>i-1,~i<0$ or
$j<0$.  The coefficients for $S_3$ and $S_4$ can be extracted from the recursion relation (\ref{recur}) as needed
to determine the $T_{nm}$ given in Tables \ref{four_coeff_tab} and \ref{five_coeff_tab}. 
It is immediately apparent that if $i=n+1$ and $j=n$ in 
Eq.\ (\ref{recur}), then
\begin{equation}
4 g_1 \sigma_{n+1,n}^{{n}}=0
\end{equation}
and so for consistency $g_1=0$, as is already known from explicit
calculation. If now in Eq.\ (\ref{recur}), we set $i=n+1$, it follows that
\begin{equation}
\sigma_{n+1,j+1}^{{n}}=\rho\left(\frac{n}{j+1}\sigma_{n,j}^{{n-1}}-\sigma_{n,j+1}^{{n-1}}\right)
\label{eq107}
\end{equation}
where $\rho=-\frac{b_3}{b_2}$. Considering values of $i$ less than
$n+1$ results in a recursion relation that requires knowing $b_4$,
$g_2$, {\it etc}.

For $j=n-1$, it follows from  (\ref{eq107}) that
\begin{equation}
\sigma_{n+1,n}^{{n}}=\rho \sigma_{n,n-1}^{{n-1}}
\end{equation}
so that
\begin{equation}
\sigma_{n+1,n}^{{n}}=\rho^n \sigma_{1,0}^{{0}}
\end{equation}
where $\sigma_{1,0}^{{1}}=b_2$. As a result, in the
expansion of  $V$, there is a contribution
\begin{equation}
V_{I}=\frac{1}{b_2}\sum_{n=0}^{\infty}\lambda^{n+1}\sigma_{n+1,n}^{{n}}\frac{L^n}{w^{n+1}}
\end{equation}
which is a geometric series whose sum is
\begin{equation}
V_{I}=\frac{\lambda}{4!}\frac{1}{w+\frac{\lambda b_3}{b_2}\log(w)}
\end{equation}

Consequently, the sum of the contributions that are of the highest
order in $L$ and $\frac{1}{w}$ at the N$^{n}$LL order of
perturbation theory gives rise to a singularity in $V$ appearing,
not when $w=0$, but rather when $w+\frac{\lambda
b_3}{b_2}\log(w)=0$. 

If now $j=n-2$ in (\ref{recur}), we find that
\begin{equation}
\sigma_{n+1,n-1}^{{n}}=\rho\left(\frac{n}{n-1}\sigma_{n,n-2}^{{n-1}}-\sigma_{n,n-1}^{{n-1}}\right)
\end{equation}
which implies
\begin{equation}
\sigma_{n+1,n-1}^{{n}}=-n\rho^n\left(\frac{1}{2}+\frac{1}{3}+\ldots+\frac{1}{n}\right)+n\rho^{n-1}\sigma_{2,0}^{{1}}
\end{equation}
where $\sigma_{2,0}^{{1}}$ is given above. Upon expressing
\begin{equation}
\frac{1}{2}+\frac{1}{3}+\ldots+\frac{1}{n}=\lim_{x\rightarrow1}\sum_{k=2}^{n}\int_{0}^{x}y^{k-1}dy   =\lim_{x\rightarrow1}\int_{0}^{x}\frac{y-y^n}{1-y}dy
\end{equation}
we now find that $V$  now has a contribution
\begin{equation}
V_{II}=\sum_{n=0}^{\infty}\lambda^{n+1}\sigma_{n+1,n-1}^{{n}}\frac{L^{n-1}}{w^{n+1}}
\end{equation}
which then becomes
\begin{equation}
V_{II}=\sum_{n=0}^{\infty}\lambda^{n+1}\frac{L^{n-1}}{w^{n+1}}\lim_{x\rightarrow1}
\left[-n\rho^n\int_{0}^{x}\frac{y-y^n}{1-y}dy+n\rho^{n-1}\sigma_{2,0}^{{1}}\right]
\end{equation}
Since
$\sum_{k=0}^{\infty}kx^{k-1}=\frac{d}{dx}\sum_{k=0}^{\infty}x^k=\frac{1}{(1-x)^2}$,
the sum and integral in the above can be evaluated in turn, leading
to
\begin{equation}
V_{II}=\frac{\lambda^2}{w\left(w+\frac{\lambda b_3}{b_2}\log w\right)}\left[\rho\left(1+\frac{\log(1-B)}{B}\right)+\sigma_{2,0}^{{1}}\right]
\end{equation}
where $B=\frac{\lambda\rho\log(w)}{w}$. Having summed to all
orders, the contributions of $\frac{L^{n-1}}{w^{n+1}}$ in the
N$^{n}$LL contribution to $V$ to obtain $V_{II}$, we again find that $V$ has a
peculiar singularity structure.

\section*{Appendix B: Massless Scalar Electrodynamics}
In this Appendix we show how the sums appearing in Eqs.\ (\ref{sqed_eq10}, \ref{sqed_eq11}) for $V_{LL}$
and $V_{NLL}$ can be evaluated in closed form by adapting the
method of characteristics \cite{Cour,Peter}. This first entails defining
\begin{equation}
w^k_{n+k} \left(\bar{g} (t), \bar{\alpha} (t), t \right) = \exp \left[ 4 \int^t_0 \gamma_1 \left(\bar{g} (\tau), \bar{\alpha} (\tau) \right) d \tau \right] p^k_{n+k} \left(\bar{g} (t), \bar{\alpha} (t) \right) \label{sqed_eq21}
\end{equation}
where
\begin{gather}
\frac{d \bar{g}(t)}{d t} = \beta^g_2 \left( \bar{g} (t), \bar{\alpha} (t) \right) \; \left( \bar{g} (0) = g \right) 
\label{sqed_eq22} 
\\
\frac{d \bar{\alpha} (t)}{d t} = \beta^\alpha_2 \left(\bar{g} (t), \bar{\alpha} (t) \right) \; \left( \bar{\alpha} (0) = \alpha \right) 
\label{sqed_eq23} 
\end{gather}
are characteristic functions.  
\footnote{Solutions to Eqs.\ (\ref{sqed_eq22}, \ref{sqed_eq23}) appear
in Ref.\ \cite{1}. They are easily obtained as $b_{1,1}^{\alpha}=0$.}
From Eqs.\ (\ref{sqed_eq21}--\ref{sqed_eq23}) it follows that
\begin{equation}
\frac{d}{dt} w^k_{n+k} (\bar{g}, \bar{\alpha}, t) = \left( \beta^g_2 (\bar{g}, \bar{\alpha}) \frac{\partial}{\partial \bar{g}} + \beta^\alpha_2 (\bar{g}, \bar{\alpha}) \frac{\partial}{\partial \bar{\alpha}}
+ 4 \gamma_1 (\bar{g}, \bar{\alpha}) \right) w^k_{n+k} (\bar{g}, \bar{\alpha}, t) 
~.
\label{sqed_eq24}
\end{equation}
Together, Eqs.\ (\ref{sqed_eq18}, \ref{sqed_eq24}) show that
\begin{equation}
w^n_{n+1} (\bar{g}, \bar{\alpha}, t) = \frac{1}{2n} \left(\beta^g_2 (\bar{g},\bar{\alpha}) \frac{\partial}{\partial \bar{g}} + \beta^\alpha_2 (\bar{g},\bar{\alpha}) \frac{\partial}{\partial \bar{\alpha}} + 4 \gamma_1 (\bar{g}, \bar{\alpha}) \right) w^{n-1}_n (\bar{g}, \bar{\alpha}, t) ~.
\label{sqed_eq25} 
\end{equation}

We now define
\begin{equation}
V_{LL}(t)= \sum^\infty_{n=0} w^n_{n+1} \left( \bar{g} (t), \bar{\alpha} (t), t \right) \bar{L}^n \phi^4
~, 
\label{sqed_eq26} 
\end{equation}
where
\begin{equation}
\bar{L} = \log \left( \frac{\phi^2}{\bar{\mu}^2 (t)} \right) 
\label{sqed_eq27} 
\end{equation}
with
\begin{equation}
\frac{d \bar{\mu} (t)}{dt} = \bar{\mu}(t) ~,~ \bar{\mu}(0)=\mu 
\label{sqed_eq28}~. 
\end{equation}
From Eqs.\ (\ref{sqed_eq10}, \ref{sqed_eq21}--\ref{sqed_eq23}, \ref{sqed_eq26}--\ref{sqed_eq28}) it follows that
\begin{equation}
V_{LL} (t = 0) = V_{LL} ~.
\label{sqed_eq29} 
\end{equation}
We see that Eqs.\ (\ref{sqed_eq24}, \ref{sqed_eq25}) lead to
\begin{equation}
w^n_{n+1} (\bar{g}, \bar{\alpha}, t) =  \frac{1}{2^nn!} \frac{d^n}{dt^n} w^0_1 (\bar{g}, \bar{\alpha}, t) \label{sqed_eq30} 
\end{equation}
so that Eq.\ (\ref{sqed_eq26}) becomes
\begin{equation}
V_{LL} (t) = \sum^\infty_{n=0} \frac{\bar{L}^n}{2^nn!} \frac{d^n}{dt^n} w^0_1 (\bar{g} (t), \bar{\alpha} (t), t) \phi^4 = w^0_1 \left(\bar{g} \left(t+ \frac{\bar{L}}{2} \right), \bar{\alpha} \left(t + \frac{\bar{L}}{2} \right), t + \frac{\bar{L}}{2} \right) 
\label{sqed_eq31}  ~.
\end{equation}
Furthermore, Eqs.\ (\ref{sqed_eq21}--\ref{sqed_eq23}, \ref{sqed_eq29}) reduce Eq.\ (\ref{sqed_eq31}) to
\begin{equation}
V_{LL} =  w^0_1 \left(\bar{g} \left(\frac{L}{2}\right), \bar{\alpha} \left(\frac{L}{2}\right), \frac{L}{2} \right)\phi^4 
\label{sqed_eq32} ~.
\end{equation}
This coincides with the result appearing in Ref.\ \cite{5}.

Having found this closed form expression for the leading-log contribution to $V(\phi)$, we turn to the next-to-leading log contribution of Eq.\ (\ref{sqed_eq11}).  The first step is to define
\begin{equation}
V_{NLL} (t) = \sum^\infty_{n=0} w^n_{n+2} \left( \bar{g} (t), \bar{\alpha} (t), t \right) \bar{L}^n \phi^4 \label{sqed_eq33} ~. 
\end{equation}
We now note that Eqs.\ (\ref{sqed_eq20}, \ref{sqed_eq24}, \ref{sqed_eq25}) together show that
\begin{equation}
\begin{split}
w^n_{n+2} &= \frac{1}{2n} \left[\left( \frac{d}{dt} w^{n-1}_{n+1} \right) + \left( \gamma_1 \left(\beta^g_2 \frac{\partial}{\partial \bar{g}} + \beta^\alpha_2 \frac{\partial}{\partial \bar{\alpha}} + 4 \gamma_1 \right) + \left(\beta^g_3 \frac{\partial}{\partial \bar{g}} + \beta^\alpha_3 \frac{\partial}{\partial \bar{\alpha}} + 4 \gamma_2 \right) \right) w^{n-1}_n \right]
\\ 
&\equiv \frac{1}{2n} \left( \frac{d}{dt} w^{n-1}_{n+1} + D (t) w^{n-1}_n \right) ~,
\end{split}
\label{sqed_eq34} 
\end{equation}
where $D(t)$ corresponds to the differential operator acting upon $w^{n-1}_n$.

Iterating Eq.\ (\ref{sqed_eq34}), we obtain
\begin{equation}
w^n_{n+2} = \frac{1}{2n} \left[ \frac{d}{dt} \left( \frac{1}{2(n-1)} \left( \frac{d}{dt} w^{n-2}_n + D (t) w^{n-2}_{n-1} \right) \right) + D (t) w^{n-1}_n \right] 
\label{sqed_eq35}~. 
\end{equation}
Repeating this $n$ times and using Eq.\ (\ref{sqed_eq30}) we find
\begin{equation}
w^n_{n+2} = \frac{1}{2^nn!} \left[ \frac{d^n}{dt^n} w^0_2 + \frac{d^{n-1}}{dt^{n-1}} D (t) w^0_1 + \frac{d^{n-2}}{dt^{n-2}} D (t) \frac{d}{dt} w^0_1 + \frac{d^{n-3}}{dt^{n-3}} D(t) \frac{d^2}{dt^2} w^0_1 + \dots + D(t) \frac{d^{n-1}}{dt^{n-1}} w^0_1 \right] ~.
\label{sqed_eq36} 
\end{equation}
The identity
\begin{equation}
\frac{d^n}{dt^n} (fg) + \frac{d^{n-1}}{dt^{n-1}} \left( f \frac{dg}{dt} \right) + \dots + \frac{d}{dt} \left( f \frac{d^{n-1} g}{dt^{n-1}} \right) + f \frac{d^n g}{dt^n} = \frac{d^{n+1}}{dt^{n+1}} \left( \phi g \right) - \phi \frac{d^{n+1}g}{dt^{n+1}}~, \; \left( \phi^\prime = f \right) 
\label{sqed_eq37} 
\end{equation}
converts Eq.\ (\ref{sqed_eq36}) to the form
\begin{equation}
w^n_{n+2} = \frac{1}{2^nn!} \left[\frac{d^n}{dt^n} w^0_2 + \frac{d^n}{dt^n} \left( \widetilde{D} w^0_1 \right) - \widetilde{D} \frac{d^n w^0_1}{dt^n} \right] 
\label{sqed_eq38} 
\end{equation}
where
\begin{equation}
\frac{d}{dt} \widetilde{D} = D~. 
\label{sqed_eq39}
\end{equation}
More explicitly, we have
\begin{gather}
 \widetilde{D}(t)w_1^0\left(\bar g(t),\bar\alpha(t),t\right)
=\left( \int\limits_{0}^t
\left[
\beta_3^g\left[\bar g(\tau),\bar\alpha(\tau) \right]\frac{\partial}{\partial \bar g(t)}
+\ldots +2\left(\gamma_1\left[  \bar g(\tau),\bar\alpha(\tau) \right]\right)^2
\right] d\tau \right)w_1^0\left(\bar g(t),\bar\alpha(t),t\right)
\label{sqed_D1}
\\
\begin{split}
\frac{d}{dt}\left[\widetilde{D}(t)w_1^0\left(\bar g(t),\bar\alpha(t),t\right)\right]=&
\left[
\beta_3^g\left[\bar g(t),\bar\alpha(t) \right]\frac{\partial}{\partial \bar g(t)}
+\ldots +2\left(\gamma_1\left[  \bar g(t),\bar\alpha(t) \right]\right)^2
\right]w_1^0\left(\bar g(t),\bar\alpha(t),t\right)
\\
&+\left( \int\limits_{0}^t
\left[
\beta_3^g\left[\bar g(\tau),\bar\alpha(\tau) \right]\frac{\partial}{\partial \bar g(t)}
+\ldots +2\left(\gamma_1\left[  \bar g(\tau),\bar\alpha(\tau) \right]\right)^2
\right] d\tau \right)
\left[\frac{d}{dt}w_1^0\left(\bar g(t),\bar\alpha(t),t\right)\right]
\end{split}
\label{sqed_D2}
\end{gather}
Eqs.~(\ref{sqed_D1}) and (\ref{sqed_D2}) ensure consistency between Eqs.~(\ref{sqed_eq36}) and (\ref{sqed_eq38}). In Eqs.~(\ref{sqed_D1}) and (\ref{sqed_D2}), $\bar g$ and $\bar\alpha$ are evaluated at $t$ when appearing in the arguments of $w_1^0$.  Derivatives  with respect to $\bar g$ and $\bar\alpha$ have
these functions evaluated at the scale $t$.  In Eq.~(\ref{sqed_D2}) , derivatives with respect to $t$ acting
on $w_1^0(\bar g(t),\bar\alpha(t),t)$ do so prior to functional derivatives 
$\partial/\partial\bar g(t)$, $\partial/\partial\bar\alpha(t)$; the last step in Eq.~(\ref{sqed_D2}) is the integral over $\tau$.

Eqs.\ (\ref{sqed_eq33}, \ref{sqed_eq38}) lead to
\begin{equation}
\begin{split}
V_{NLL} (t) = w^0_2 \left( \bar{g} ( t + \frac{\bar{L}}{2} ), \bar{\alpha} ( t + \frac{\bar{L}}{2} ), t 
+ \frac{\bar{L}}{2} \right) 
&+ \widetilde{D} \left( t + \frac{\bar{L}}{2} \right) w^0_1 \left( \bar{g} ( t + \frac{\bar{L}}{2} ), \bar{\alpha} ( t + \frac{\bar{L}}{2} ), t + \frac{\bar{L}}{2} \right)
\\
 &- \widetilde{D} (t) w^0_1 \left( \bar{g} ( t + \frac{\bar{L}}{2} ), \bar{\alpha} ( t + \frac{\bar{L}}{2} ), t + \frac{\bar{L}}{2} \right) ~.
\end{split}
\label{sqed_eq40} 
\end{equation}
It is evident that
\begin{equation}
V_{NLL} (t=0) = V_{NLL} 
\label{sqed_eq41}
\end{equation}
and so $V_{NLL}$ of Eq.\ (\ref{sqed_eq11}) is
\begin{equation}
V_{NLL} = \left[w^0_2 \left( \bar{g} ( \frac{L}{2} ), \bar{\alpha} (\frac{L}{2}), \frac{L}{2} \right) + \widetilde{D} (\frac{L}{2}) w^0_1 \left(\bar{g} (\frac{L}{2}), \bar{\alpha} (\frac{L}{2}), \frac{L}{2} \right) \right] 
\label{sqed_eq42} 
\end{equation}
(Since $\widetilde{D}^\prime (t) = D (t)$, we can set $\widetilde{D}(0)=0$.)\\

As a result, we see that if $\beta^g_2, \beta^g_3, \beta^\alpha_2, \beta^\alpha_3, \gamma_1, \gamma_2, p^0_1$ and $p^0_2$ are known, then $V_{NLL}$ is fully determined.  It is seen that this approach can also be used to find closed form expressions for $V_{N^PLL} (p \geq 2)$.


\begin{thebibliography}{99}
\bibitem{1}S.\ Coleman and E.\ Weinberg, Phys.\ Rev.\ D7 (1973) 1888.
\bibitem{2}S.\ Weinberg Phys.\ Rev.\ D7 (1973) 2887.
\bibitem{3}R.\ Jackiw, Phys.\ Rev.\ D9 (1974) 1686.
\bibitem{4}M.\ Sher, Phys.\ Rep.\ 179 (1989) 273.
\bibitem{5}V.\ Elias, R. B. Mann,  D. G. C.\ McKeon and T. G. Steele, Phys.\ Rev.\ Lett. 91  (2003) 251601;\\
V.\ Elias, R.B.\ Mann, D.G.C.\ McKeon, T.G.\ Steele, Nucl.\ Phys.\ B678 (2004) 147; Erratum-ibid B 703 (2004) 413.
\bibitem{6}G.\ 't Hooft and M.\ Veltman, Nucl.\ Phys.\ B44 (1972) 189.
\bibitem{7}C.\ Ford and D.R.T.\ Jones, Phys. Lett. B 274 (1992) 409; ibid 285 (1992) 409.
\bibitem{10}Hamad Alhendi, Phys.\ Rev.\ D37 (1988) 3749.
\bibitem{ndili} F.N.\ Ndili, arXiv:0708.0836.
\bibitem{multi_scale_einhorn} M.B.\ Einhorn and D.R.T.\ Jones, Nucl.\ Phys.\ B230 [FS10] (1984) 261.
\bibitem{multi_scale_ford} C.\ Ford and C.\ Wiesendanger, Phys.\ Lett.\ B398 (1997) 342;\\
C.\ Ford and C.\ Wiesendanger, Phys.\ Rev.\ D55 (1997) 2202.
\bibitem{vic} V.~Elias and D.G.C. McKeon, Can.\ J.\ Phys.\ 81 (2006) 131.
\bibitem{8}B.\ Kastening, Phys.\ Lett.\ B283 (1992) 287.
\bibitem{9}H.\ Kleinert, J.\ Neu, V.\ Schulte-Frohlinde, K.G.\ Chetyrkin, S.A.\ Larin, Phys. Lett. B 272 (1991) 39; ibid. 319 (1993) 545. 
\bibitem{salam} A.\ Salam and J.\ Strathdee, Phys.\ Rev.\ D9 (1974) 1129.
\bibitem{more_gerry} A.\ Kotikov and D.G.C.\ McKeon, Can.\ J.\ Phys.\ 72 (1994) 250.
\bibitem{11} C.\ Ford, I.\ Jack, D.R.T.\ Jones, Nucl.\ Phys.\ B387 (1992) 373; Erratum-ibid. B504 (1997) 551.
\bibitem{Cour} R. Courant and D. Hilbert, ``Methods of Mathematical Physics Vol. II" (Interscience N.Y.) 1966 Chap. 11.
\bibitem{Peter} A. Peterman, Phys.\ Rep.\ 53C (1979) 157.


\end{thebibliography}
\end{document}